\documentstyle[aps,preprint,tighten,eqsecnum,epsfig,floats,prd]{revtex}

\newcommand{\rf}[4]{#1 {\bf #2}, #3 (#4)}

\newcommand{\pr}{Phys.\ Rev.}

\newcommand{\physl}{Phys.\ Lett.}

\newcommand{\np}{Nucl.\ Phys.}
\newcommand{\npps}[3]{\emph{Nucl.\ Phys.} {\bf B} (Proc.\ Suppl.) #1, 
	#2 (#3)}
\newcommand{\cm}{Commun.\ Math.\ Phys.}
\newcommand{\beq}{\begin{equation}}
\newcommand{\eeq}{\end{equation}}
\newcommand{\beqa}{\begin{eqnarray}}
\newcommand{\eeqa}{\end{eqnarray}}

\newcommand{\adj}{\dagger}
\newcommand{\oa}[1]{${\cal O}(a^{#1})$}

\newcommand{\eqref}[1]{(\ref{#1})}

\newcommand{\tr}{\text{Tr}}

\newcommand{\qhat}{\hat{q}}
\newcommand{\muhat}{\hat{\mu}}
\newcommand{\nuhat}{\hat{\nu}}

\begin{document}

\preprint{\vbox{
\rightline{ADP-01-01/T436}
\rightline{FSU-CSIT-01-02}}}

%\draft

\title{Infinite Volume and Continuum Limits of the Landau-Gauge Gluon
       Propagator}

\author{Fr\'ed\'eric D.R.\ Bonnet$^a$, Patrick O.\ Bowman$^{a,b}$,
        Derek B.\ Leinweber$^a$, \\ 
        Anthony G.\ Williams$^a$ and James M.\ Zanotti$^{a}$}
\address{$^a$ Special Research Centre for the Subatomic Structure of Matter 
and \\ The Department of Physics and Mathematical Physics,
Adelaide University, Australia 5005}
\address{$^b$ Department of Physics and 
School for Computational Science and
Information Technology, Florida State University, Tallahassee FL 32306, USA}

\date{\today}

\maketitle

\begin{abstract} 
We extend a previous improved action study of the Landau gauge gluon
propagator, by using a variety of lattices with spacings from $a = 0.17$
to 0.41 fm, to more fully explore finite volume and discretization
effects.  We also extend a previously used technique for minimizing
lattice artifacts, the appropriate choice of momentum variable or
``kinematic correction'', by considering it more generally as a
``tree-level correction''.  We demonstrate that by using tree-level
correction, determined by the tree-level behavior of the action being
considered, it is possible to obtain scaling behavior over a very wide
range of momenta and lattice spacings.  This makes it possible to
explore the infinite volume and continuum limits of the Landau-gauge
gluon propagator.

%This method can be extended to any choice of gauge action and to
%studies of the momentum-dependence of other Green's functions.
\end{abstract}

% PACS 12.38.Gc, 11.15.Ha, 12.38.Aw, 14.70.Dj 

%\vspace{1cm}

\section{Introduction}

There has long been interest in the infrared behavior of the 
gluon propagator as a probe into the mechanism of confinement~\cite{Man99} and  
lattice studies focusing on its ultraviolet behavior have been used to 
calculate the running QCD coupling~\cite{Bec99}.  In this report we use the 
propagator as a test-bed for an improved action and also as a means to 
investigate a general tree-level correction technique.

The infrared part of any lattice calculation may be affected by the
finite volume of the lattice.  Larger volumes mean either more lattice
points (with increased computational cost) or coarser lattices (with
corresponding discretization errors).  Improved actions have been
shown to be effective at reducing discretization errors at a given
lattice spacing in studies of the static quark potential~\cite{Lep96}
and the hadron spectrum~\cite{Lee99} and have become a
necessary part of finite temperature studies~\cite{introT}.  The
desire for larger physical volumes thus provides strong motivation for
using improved actions.  We study the gluon propagator, in Landau
gauge, in quenched QCD (pure SU(3) Yang-Mills), using the mean-field
(tadpole) improved~\cite{Lep93} version of the tree-level, \oa{2}
Symanzik improved gauge action~\cite{Sym83,Wei83,Lus85}.

To assess the effects of finite lattice spacing, we calculate the
propagator on a set of lattices from $8^3\times 16$ at $\beta = 3.75$
having $a = 0.413$ fm to $16^3 \times 32$ at $\beta = 4.38$ having $a
= 0.167$ fm.  To assist us in observing possible finite volume
effects, we add to this set a $16^3 \times 32$ lattice at $\beta =
3.92$ with $a = 0.353$, which has the very large physical size of
$5.65^3 \times 11.30 \text{ fm}^4$.  Some preliminary results of this
work were reported in Ref.~\cite{Bon00}.

We will show that tree-level correction reduces rotational symmetry
breaking and dramatically improves the ultraviolet behavior of the
propagator and thus the approach to the continuum limit.  For lattices
as coarse as 0.17 fm the gluon propagator has surprisingly good
behavior for the entire range of available momenta.  The infrared
behavior of the gluon propagator is robust even with an extremely
coarse lattice spacing of 0.41 fm.  Our calculations on a lattice with
a large volume indicates that finite volume effects are small.  The
Landau gauge gluon propagator is again found to be infrared finite,
in agreement with earlier studies.  The combination of an improved
action with appropriate tree-level correction appears to be a powerful
tool.  The generalization of these methods to the study of other
Green's functions will be discussed in a forthcoming
work~\cite{treeimp}.

\section{The Landau gauge gluon propagator}

We employ the tree-level, mean-field improved gauge action of L\"{u}scher 
and Weisz~\cite{Wei83,Lus85}
\beqa
S_{\text{l}} & = & \frac{5\beta}{3N_c} \sum_{\text{pl}} \tr \Bigl\{
	1 - \frac{1}{2} \bigl( P_{\mu\nu} + P_{\mu\nu}^\adj \bigr) \Bigr\} 
    - \frac{\beta}{12N_c u_0^2} \sum_{\text{rect}} \tr \Bigl\{
	1 - \frac{1}{2} \bigl( R_{\mu\nu} + R_{\mu\nu}^\adj \bigr) \Bigr\}
      \nonumber \\  
   & = & S_{\text{cont}}+ {\cal O}(a^4) + {\cal O}(a^2g^2),
\label{eq:S_TILW}
\eeqa
where $P_{\mu\nu}$ and $R_{\mu\nu}$ are the usual plaquette and rectangle 
operators
\beq
P_{\mu\nu}(x) = U_\mu(x) U_\nu(x+\muhat) U_\mu^\dagger(x+\nuhat) 
			U_\nu^\dagger(x)
\eeq
and
\beqa
R_{\mu\nu}(x) & = & U_{\mu}(x)U_{\nu}(x+\hat{\mu})
		U_{\nu}(x+\hat{\nu}+\hat{\mu})		
U^{\dagger}_{\mu}(x+2\hat{\nu})U^{\dagger}_{\nu}(x+\hat{\nu})
		U^\dagger_{\nu}(x) \nonumber \\
	 &  & +  U_{\mu}(x)U_{\mu}(x+\hat{\mu})U_{\nu}(x+2\hat{\mu})
		U^{\dagger}_{\mu}(x+\hat{\mu}+\hat{\nu})
		U^{\dagger}_{\mu}(x+\hat{\nu})U^\dagger_{\nu}(x),
\eeqa
and $N_c = 3$ is the number of colors.
We use the plaquette definition for the tadpole factor 
\begin{equation}
u_0=\left(\frac{1}{N_c}{\cal R}e{\rm Tr}\langle P_{\mu\nu}\rangle
\right)^{\frac{1}{4}}.
\end{equation}
Our gauge field configurations were generated using the 
Cabbibo-Marinari~\cite{Cab82} pseudo-heatbath algorithm with
appropriate link partitioning \cite{LinkPart}.

Given that the gauge links $U_\mu(x)$ are expressed in terms of the continuum
gluon fields as
\begin{equation}
U_\mu(x) = {\cal P} e^{ig\int_0^1A_\mu(x+at\muhat)dt},
\end{equation}
the dimensionless lattice gluon field $A_{\mu}(x)$ may be obtained from
\begin{equation}
A_\mu(x+\muhat/2) = \frac{1}{2igu_0} \bigl\{ U_\mu(x)-U_\mu^\adj(x) 
   \bigr\}_{\text{traceless}}
\label{eq:gluonField}
\end{equation}
which is accurate to ${\cal O}(a^2)$.  This is, of course, only one of many 
possible ways to calculate the gluon field on the lattice.  In 
Eq.~\eqref{eq:gluonField}, $A_\mu$ is calculated at the midpoint of the link
to remove ${\cal O}(a)$ terms.  Note that we have also included the tadpole 
factor to improve the normalization.

We calculate the gluon propagator in coordinate space
\begin{equation}
D_{\mu\nu}^{ab}(x,y) \equiv \langle \, A_\mu^a(x) \, A_\nu^b(y) \,
\rangle \, , 
\end{equation}
using Eq.~\eqref{eq:gluonField}.  To improve statistics, we use translational
invariance and calculate
\begin{equation}
D_{\mu\nu}^{ab}(y) = \frac{1}{V} \bigl\langle \sum_x A_\mu^a(x)
   A_\nu^b(x + y) \bigr\rangle.
\end{equation} 
The quantity that will be of interest to us is the scalar part of the 
propagator in momentum space, so first we take the trace over color components
\begin{equation}
D_{\mu\nu}(y) = \frac{1}{N_c^2-1} \sum_a D_{\mu\nu}^{aa}(y),
\end{equation}
then sum over the Lorentz components\footnote{The Landau gauge condition in
momentum space, $q_\mu D_{\mu\nu}(q) = 0$ places a constraint on the Lorentz
components of the propagator so that, for non-zero momentum, there are $N_d-1$
degrees of freedom~\cite{Cuc99b}.}
of the Fourier transform
\begin{equation}
D(\qhat) = \frac{1}{N_d-1} \sum_{\mu} \sum_y e^{i \qhat \cdot y} D_{\mu\mu}(y)
\end{equation}
and 
\begin{equation}
D(0) = \frac{1}{N_d} \sum_{\mu} \sum_y  D_{\mu\mu}(y). 
\end{equation}
$N_d$ is the number of space-time dimensions and the available
momentum values, $\qhat$, are given by
\begin{equation}
\qhat_\mu  = \frac{2 \pi n_\mu}{a L_\mu}, \qquad
n_\mu \in  \Bigl( -\frac{L_\mu}{2}, \frac{L_\mu}{2} \Bigr].
\label{eq:qhat}
\end{equation}
The range of $\qhat$ is determined by the fact that our lattices have an
even number of points in each direction and that we use periodic boundary
conditions.
In the continuum, the scalar propagator is related to the full 
propagator by
\begin{equation}
D_{\mu\nu}^{ab}(q) =
\bigl( \delta_{\mu\nu}-\frac{q_{\mu}q_{\nu}}{q^2} \bigl) \delta^{ab}D(q^2)
\label{eq:landau_prop}
\end{equation}
in Landau gauge.

Landau gauge is a smooth gauge that preserves the Lorentz invariance of the 
theory, so it is a popular choice.  We work in Landau gauge for ease of 
comparison with other studies, and because it is the simplest covariant gauge to 
implement on the lattice.  All configurations were gauge fixed by maximizing 
an \oa{2} improved Landau gauge fixing functional using Conjugate Gradient 
Fourier Acceleration~\cite{Cuc98} as described in Ref.~\cite{Bon99}.

\section{Tree-level Correction}

One thing that is known about the gluon propagator is its perturbative, 
asymptotic behavior.  In the spirit of improvement, we can use this knowledge to 
augment our lattice results and make better contact with the continuum.  In the 
continuum, as $p^2\rightarrow\infty$, the propagator has the form
\begin{equation}
D(p) = \frac{1}{p^2}
\end{equation}
up to logarithmic corrections.  A well known artifact of the lattice is that for 
a free massless boson with an unimproved action the lattice propagator has the 
form 
\begin{equation}
D(\qhat) = \frac{1}{\frac{4}{a^2}\sum_{\mu}\sin^2 
	\bigl( \frac{\qhat_\mu a}{2} \big)}.
\label{eq:wil_tree}
\end{equation}
It has been argued, in Ref.~\cite{Lei98} and elsewhere, that the correct 
momentum variable to use when examining the gluon propagator on the lattice, 
with the Wilson action, is not Eq.~\eqref{eq:qhat}, 
but\footnote{Many authors have $q$ and $\qhat$ defined the other way around,
but in this context our terminology is more natural.}
\begin{equation}
q_{\mu} \equiv \frac{2}{a} \sin\frac{\qhat_{\mu} a}{2}.
\end{equation}
It has been observed that this choice ensures that the propagator takes its 
asymptotic form at large lattice momenta~\cite{Lei98,Mar95}.

The improved action Eq.~\eqref{eq:S_TILW} together with the gluon
field defined in Eq.~(\ref{eq:gluonField}) has the \oa{2} improved tree-level 
behavior~\cite{Sym83,Wei83}
\begin{equation}
D^{-1}(\qhat) = \frac{4}{a^2}\sum_{\mu}\biggl\{ \sin^2 
	\Bigl( \frac{\qhat_\mu a}{2} \Bigr)
	+ \frac{1}{3}\sin^4 \Bigl( \frac{\qhat_\mu a}{2} \Bigr) 
	\biggr\},
\label{eq:imp_tree}
\end{equation}
and we will use Eqs.~\eqref{eq:wil_tree} and~\eqref{eq:imp_tree} to obtain 
the correct momentum variable for each action. 
To emphasize the nonperturbative aspects of the propagator, we divide it by
its perturbative, $1/q^2$ result. 
Hence, all figures are plotted against $q^2 D(q^2)$, which is expected to 
approach a constant up to logarithmic corrections as $q^2\to\infty$.  We will 
see that this also makes for a stringent test of the ultraviolet behavior of 
the propagator.
We will work with the momentum variables defined as
\begin{equation}
q_{\mu}^W \equiv \frac{2}{a} \sin\frac{\qhat_{\mu} a}{2},
\label{eq:wil_momenta}
\end{equation}
and
\begin{equation}
q_\mu^I \equiv \frac{2}{a}\sqrt{ \sin^2 
	\Bigl( \frac{\qhat_\mu a}{2} \Bigr)
	+ \frac{1}{3}\sin^4 \Bigl( \frac{\qhat_\mu a}{2} \Bigr) 
	} \, ,
\label{eq:imp_momenta}
\end{equation}
for the Wilson and improved actions respectively.  A similar momentum variable 
was used in the study of the gluon propagator in Ref.~\cite{Ma00}.

In the language of continuum physics
\begin{equation}
\label{eq:continuum}
p^2 D(p^2) = \frac{1}{1+\Pi(p^2)} = \frac{D(p^2)}{D^{\text{tree}}(p^2)}
\end{equation}
where $\Pi(p^2)$ is the scalar vacuum polarization.  In the asymptotic region,
$1 / [1+\Pi(p^2)] \rightarrow 1$ up to logarithmic corrections.  We argue 
that it is the lattice version of $D(p^2) / D^{\text{tree}}(p^2)$ that
will most rapidly approach its continuum form as the lattice spacing is
reduced and we will later graphically demonstrate this.  The essential
point is that at large momenta the lattice gluon propagator will
experience asymptotic freedom just as in the continuum, i.e., the
ultraviolet propagator will approach its tree-level form.
Thus on the lattice we expect to find $D(p^2) / D^{\text{tree}}(p^2)\to 1$
for large $p^2$ even though the ultraviolet lattice artifacts in both
$D(p^2)$ and $D^{\text{tree}}(p^2)$ may themselves be large.
We will refer to this procedure for minimizing ultraviolet
lattice artifacts as \emph{tree-level correction}.  This philosophy is
similar to that applied in recent studies of the quark
propagator~\cite{Sku00}.
In figures where $q^2D(q^2)$ is plotted vs.\ $q$, 
the ``$q$'' in $q^2D(q^2)$ (plotted on the y-axis) is always the same as
the $q$ that is used on the x-axis, where $q=\qhat$, $q^W$ or $q^I$
as described in the text.

The bare, dimensionless lattice gluon propagator $D(qa)$ is related to the
renormalized continuum propagator $D_R(q;\mu)$ by
\begin{equation}
\label{eq:renorm}
a^2 D(qa) = Z_3(\mu,a) D_R(q;\mu),
\end{equation}
for momenta, $q$, sufficiently small compared to the cutoff, $\pi/a$.
$D_R(q;\mu)$ is independent of $a$ for sufficiently fine lattices;
i.e.\ in the scaling regime.
The renormalization constant $Z_3(\mu,a)$ is determined by imposing a 
renormalization condition at some chosen renormalization scale $\mu$, e.g.,
\begin{equation}
\label{eq:b.c.}
D_R(q)|_{q^2=\mu^2} = \frac{1}{\mu^2}.
\end{equation}
The renormalized gluon propagator can be computed both nonperturbatively on 
the lattice and perturbatively in the continuum for choices of the 
renormalization point in the ultraviolet.  It can then be related to the 
propagator in other continuum renormalization schemes such as 
$\overline{\text{MS}}$.

\section{Results}
\label{sec:results}

\subsection{Analysis Overview}

The gluon propagator has been calculated on seven different lattices, the 
details of which are listed in Table~I.  Note that the first 
two are labeled ``1w'' and ``1i''.  These have the same number of lattice 
points at almost the same spacing (hence approximately the same physical 
volume), but 1w was generated with the standard, Wilson gauge action, 
while 1i used the \oa{2} improved action~\eqref{eq:S_TILW}.
Lattice 6 was generated with the Wilson action and used to study the gluon
propagator in Ref.~\cite{Lei98}.  A value for the tadpole factor has been
obtained for $\beta=6.0$ of $u_0 = 0.878$ and this has been used to normalize
the propagator with respect to the other lattices.  It will be used here for 
comparison purposes as it is finer than the other lattices.
Configurations on lattices 2--5 were generated with the \oa{2} improved action.
All of the propagators are plotted in physical units, where the scale has been 
determined by the static quark potential with a string tension of 
$\sqrt{\sigma} = 440$ MeV.  Details of this calculation may be found in 
Ref.~\cite{spacing}.

\begin{table}[bt]
\label{table:latlist}
\centering
\begin{tabular}{cccccc}
   & Dimensions	     & $\beta$ & $a$ (fm) & Volume $\text{(fm}^4\text{)}$ &
Configurations \\
\hline
1w & $16^3\times 32$ &   5.70  &   0.179  & $2.87^3 \times 5.73$  & 100 \\
1i & $16^3\times 32$ &   4.38  &   0.166  & $2.64^3 \times 5.28$  & 100 \\
2  & $10^3\times 20$ &   3.92  &   0.353  & $3.53^3 \times 7.06$  & 100 \\
3  & $8^3 \times 16$ &   3.75  &   0.413  & $3.30^3 \times 6.60$  & 100 \\
4  & $16^3\times 32$ &   3.92  &   0.353  & $5.65^3 \times 11.30$ & 100 \\
5  & $12^3\times 24$ &   4.10  &   0.270  & $3.24^3 \times 6.48$  & 100 \\
6  & $32^3\times 64$ &   6.00  &   0.099  & $3.18^3 \times 6.34$  &  75 \\
\end{tabular}
\caption{Details of the lattices used to calculate the gluon propagator.  
Lattices 1w and 1i have the same dimensions and approximately the same lattice
spacing, but were generated with the Wilson and improved actions respectively.
Lattice 6 was generated with the Wilson action.}
\end{table}

Data points that come from momenta lying entirely along a spatial
Cartesian direction are indicated with a square while points from
momenta entirely in the temporal direction are marked with a triangle.  As
the time direction is longer than the spatial ones any difference
between squares and triangles may indicate that the propagator is
affected by the finite volume of the lattice.  Data points from
momenta entirely on the four-diagonal are marked with a diamond.
Systematic separation of data points taken on the diagonal from those
in other directions indicates violation of rotational symmetry.

In the continuum, the scalar function is rotationally
invariant.  Although the hypercubic lattice breaks O(4) invariance, it does 
preserve the subgroup of discrete rotations Z(4).  In our case, this symmetry
is reduced to Z(3) as one dimension will be twice as long as the other three
in each of the cases studied.  We exploit this discrete rotational symmetry to 
improve statistics through Z(3) averaging.  This is best explained through a 
simple example.  Consider
the propagator at momentum $q=(3,2,1,4)$ (say).  Z(3) symmetry means that
\beq
D(3,2,1,4) = D(2,3,1,4) = D(2,1,3,4) 
   = D(1,2,3,4) = D(1,3,2,4) = D(3,1,2,4) 
\eeq
so we calculate the propagator for each of these values of momentum, and then
average the results.

\subsection{Tree-Level Correction and Rotational Symmetry}

\begin{figure}[p]
\begin{center}
\epsfig{figure=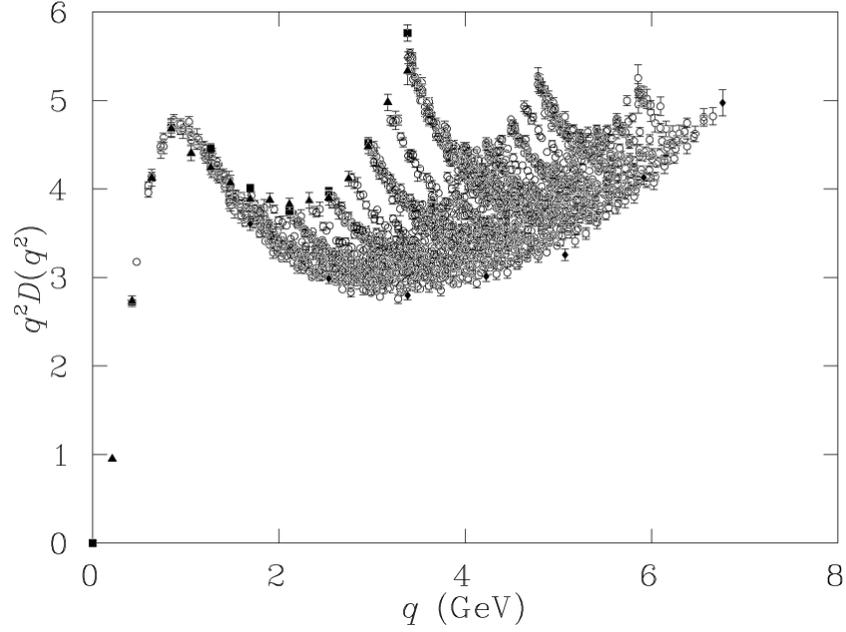,angle=90,width=11cm}
\end{center}
\caption{Uncorrected gluon propagator from lattice 1w ($\beta=5.70,16^3 \times 
32$, Wilson action), plotted as a function of $\qhat$.  The dramatic ``fanning'' 
is caused by finite spacing errors which quickly destroy the signal at large 
momenta. }
\label{fig:b570QhatAll}
\end{figure}

\begin{figure}[p]
\begin{center}
\epsfig{figure=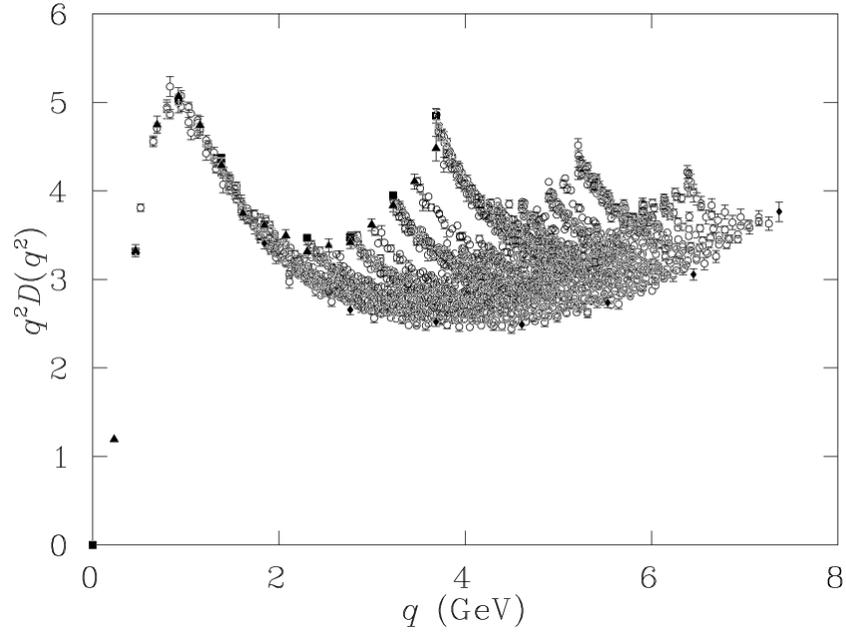,angle=90,width=11cm}
\end{center}
\caption{Uncorrected gluon propagator from lattice 1i ($\beta=4.38,16^3 \times 
32$, improved action), plotted as a function of $\qhat$.  Lattice artifacts are 
reduced by the improved action, but are still large.}
\label{fig:b438QhatAll}
\end{figure}

The ``raw'' gluon propagator from lattices 1w and 1i is shown in
Figures~\ref{fig:b570QhatAll} and~\ref{fig:b438QhatAll} respectively.
Both of these have been plotted as functions of $\qhat$,
Eq.~\eqref{eq:qhat}, for all available momenta, and both show severe
ultraviolet noise.  We may take some comfort from the observation that
the signal degradation is not as bad in the improved case where the
finite spacing errors do not exceed the infrared peak and the UV tail
is generally flatter.  However, neither result looks at all
satisfactory at large momenta. No data cuts or tree-level correction
have yet been used.

\begin{figure}[p]
\begin{center}
\epsfig{figure=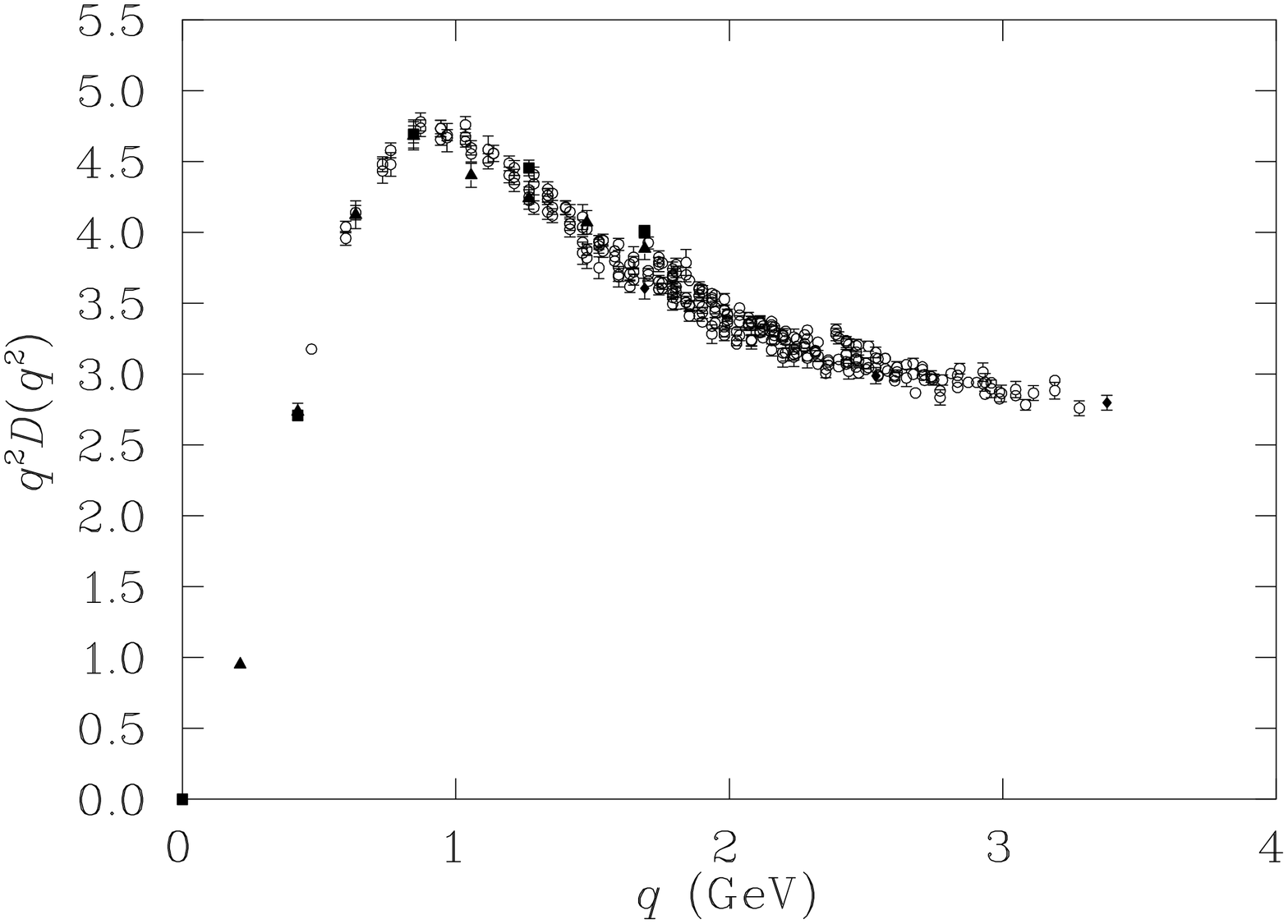,angle=90,width=11cm}
\end{center}
\caption{Uncorrected gluon propagator from lattice 1w ($\beta=5.70,16^3 \times 
32$, Wilson action), plotted as a function of $\qhat$ with the momentum 
``half-cut'' applied.}
\label{fig:b570QhatCut}
\end{figure}

\begin{figure}[p]
\begin{center}
\epsfig{figure=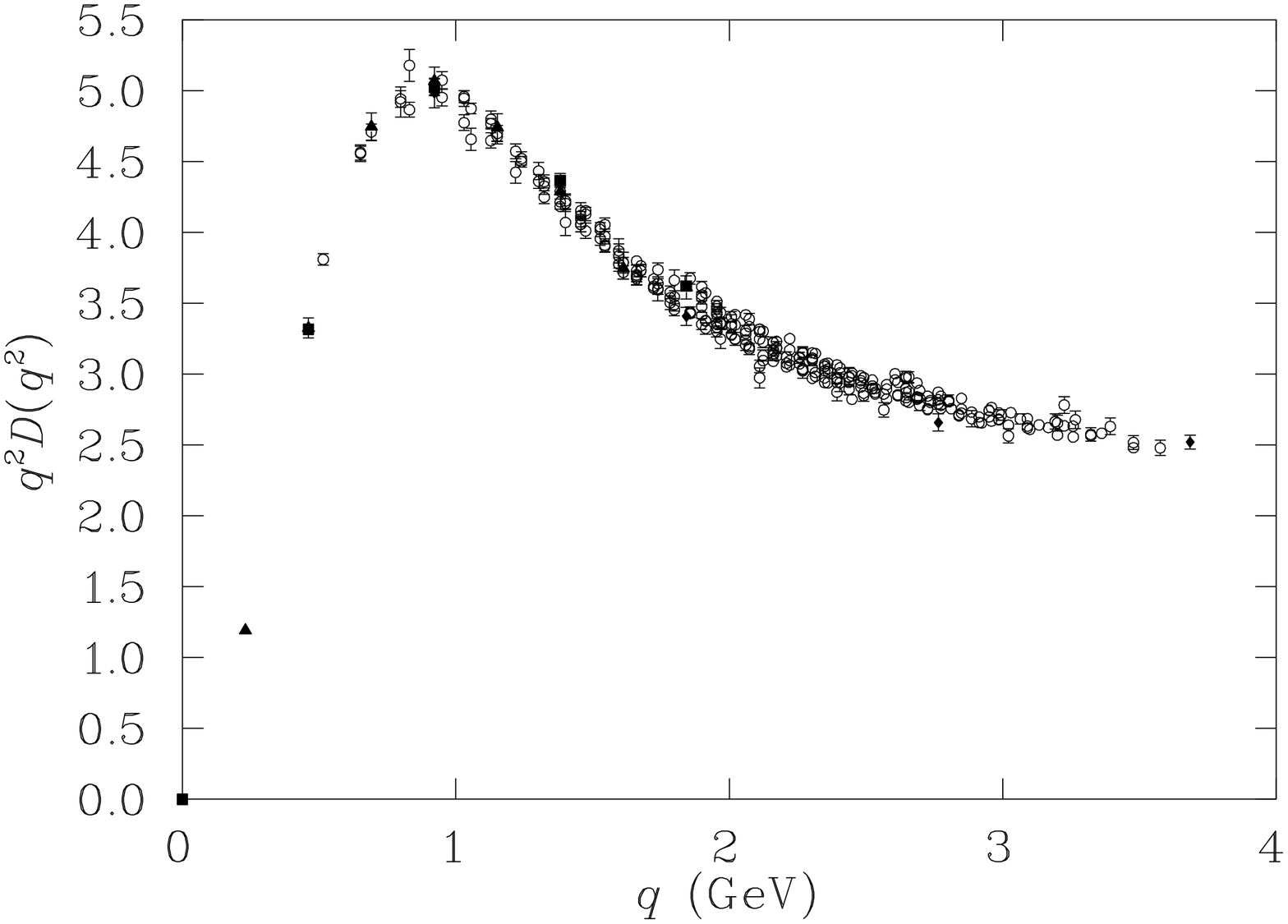,angle=90,width=11cm}
\end{center}
\caption{Uncorrected gluon propagator from lattice 1i
($\beta=4.38,16^3 \times 32$, improved action), plotted as a function
of $\qhat$ with the momentum ``half-cut'' applied.  The improved
propagator has different normalization to the Wilson case due to a
difference in the $Z_3$ renormalization constant.}
\label{fig:b438QhatCut}
\end{figure}

The most obvious way to deal with this noise is to apply an ultraviolet
cut, considering only momenta out to half of the Brillouin zone.  For each
of the four Cartesian directions,
\begin{equation}
\qhat \le \frac{\pi}{2a}.
\end{equation}
We refer to this as the ``half-cut'' and in Fig.~\ref{fig:b570QhatCut}
and Fig.~\ref{fig:b438QhatCut} we see that this removes the worst of
the artifacts.  The two propagators, show plausible asymptotic
behaviors, but there are still clear signs of lattice artifacts and we
have lost a lot of data in the ultraviolet.  While neither of these
shortcomings is a significant problem for studies of the infrared,
we will show that something as crude as the half-cut is not necessary
and we can do much better at minimizing lattice artifacts.

\begin{figure}[p]
\begin{center}
\epsfig{figure=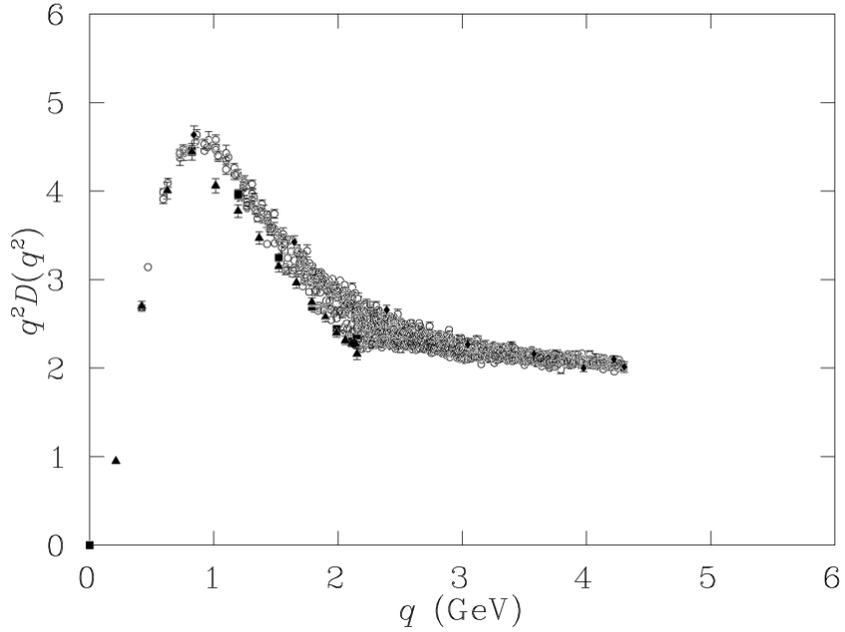,angle=90,width=11cm}
\end{center}
\caption{Uncut gluon propagator from lattice 1w ($\beta=5.70,16^3 \times 32$, 
Wilson action), plotted as a function of $q^W$ for all momenta.
The tree-level correction has greatly reduced discretization errors
from those seen in Fig.~\protect\ref{fig:b570QhatAll}. }
\label{fig:b570PhysAll}
\end{figure}

\begin{figure}[p]
\begin{center}
\epsfig{figure=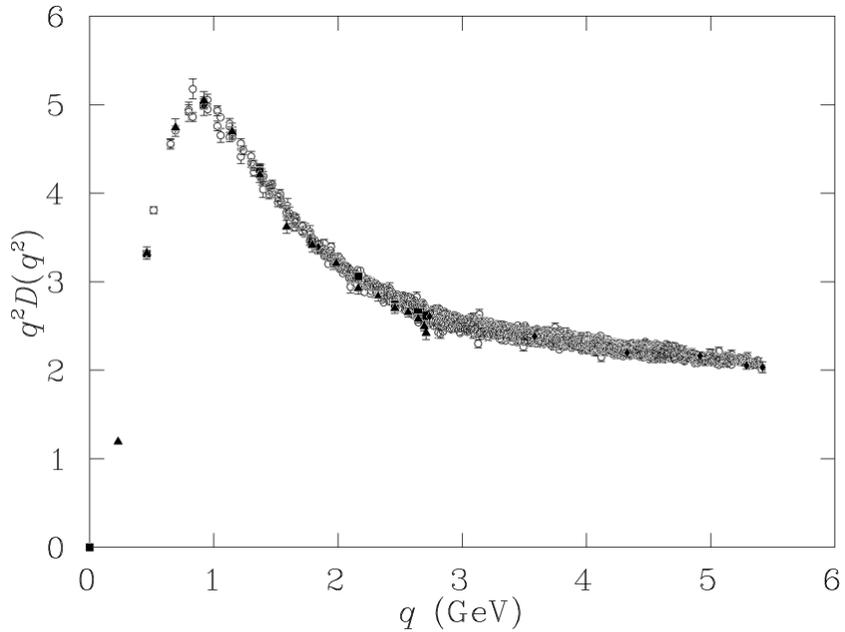,angle=90,width=11cm}
\end{center}
\caption{Uncut gluon propagator from lattice 1i ($\beta=4.38,16^3
\times 32$, improved action), plotted as a function of $q^I$ for all
momenta.  The combination of improved action and tree-level correction
has produced a remarkably clean signal over the entire range of
accessible momenta.  This figure should be compared with
Fig.~\protect\ref{fig:b438QhatAll}, and with
Fig.~\protect\ref{fig:b570PhysAll} for the Wilson action at a similar
lattice spacing. }
\label{fig:b438PhysAll}
\end{figure}

We have already argued the case for applying a tree-level correction through
the use of the alternative momentum variables derived from the tree-level 
behavior of the actions.  The effect of doing this is 
seen in Fig.~\ref{fig:b570PhysAll} and Fig.~\ref{fig:b438PhysAll}, where the 
Wilson propagator has been plotted as a function of $q^W$ 
($q^W D(q^W)$ vs. $q^W$) and the improved 
propagator as a function of $q^I$ ($q^I D(q^I)$ vs. $q^I$) for all momenta of 
the Brillouin zone.  
Comparing these to 
Figs.~\ref{fig:b570QhatAll} and~\ref{fig:b438QhatAll}, we 
see an excellent restoration of rotational symmetry all the way to the edge of 
the Brillouin zone.  This is especially true of the improved action case in 
Fig.~\ref{fig:b438PhysAll}. 
The propagators also appear to be approaching their 
asymptotic, perturbative values.  Later, momentum cuts will be applied to the 
data to further eliminate lattice artifacts, but for the moment it is 
interesting to keep all data, as they provide insight into the behavior of 
lattice simulations.

Both Figs.~\ref{fig:b570PhysAll} and~\ref{fig:b438PhysAll} are consistent 
with the study of Ref.~\cite{Lei98}, but the discrepancy between 
diagonal and Cartesian points in 
Fig.~\ref{fig:b570PhysAll} is a clear sign of rotational symmetry breaking
in the unimproved case.
With the Wilson action, the quality of the data is suffering from the 
coarseness of the lattice.  As we might hope, the improved propagator in 
Fig.~\ref{fig:b438PhysAll} shows excellent agreement between diagonal and 
Cartesian points, and the data is generally less spread.  The propagator from 
the improved action has better rotational symmetry at the same lattice spacing. 
 Less easy to understand is the slight suppression of the temporal points
(triangles)  in the Wilson case, 
Fig.~\ref{fig:b570PhysAll}.  The time axis of this lattice (as with all the 
lattices considered here) is twice as long as the other three axes, so 
different values for the points along the long axis would normally be 
interpreted as a finite volume effect, yet there is no sign of it in the 
improved case (which has 
approximately the same physical volume).  There is a difference between the
improved and unimproved cases in the amplitudes of the propagators, but this
is accounted for by renormalization and will be discussed below.

%\begin{figure}[t]
%\begin{center}
%\epsfig{figure=b438PhysQ.ps,angle=90,width=11cm}
%\end{center}
%\caption{Gluon propagator from lattice 1i (improved action), plotted as a
%function of $q^W$.  The fanning of the data and poor UV behavior demonstrate
%that this is a poor choice of momentum variable for this action, i.e., it
%corresponds to an incorrect tree-level correction.}
%\label{fig:b438PhysQ}
%\end{figure}

Out of curiosity the gluon propagator from lattice 1i has
also been examined 
as a function of $q^W$, which we have already argued to be inappropriate.  Not 
surprisingly, this leads to a ``propagator'' that suffers badly from lattice 
artifacts.  We have not included a figure here, but the resulting
propagator droops strongly in the ultraviolet.  This is clearly a poor
choice of momentum variable for this action as expected on the basis of
our tree-level correction.  For best 
results at finite lattice spacing, the correct momentum variable is determined 
by the appropriate tree-level behavior, which in turn is defined
by the choice of action and gluon field definition.
For the rest of this report it shall be implicit that when discussing 
quantities from the Wilson action, $q^W$ is used, and $q^I$ is used with
the improved action.

\begin{figure}[p]
\begin{center}
\epsfig{figure=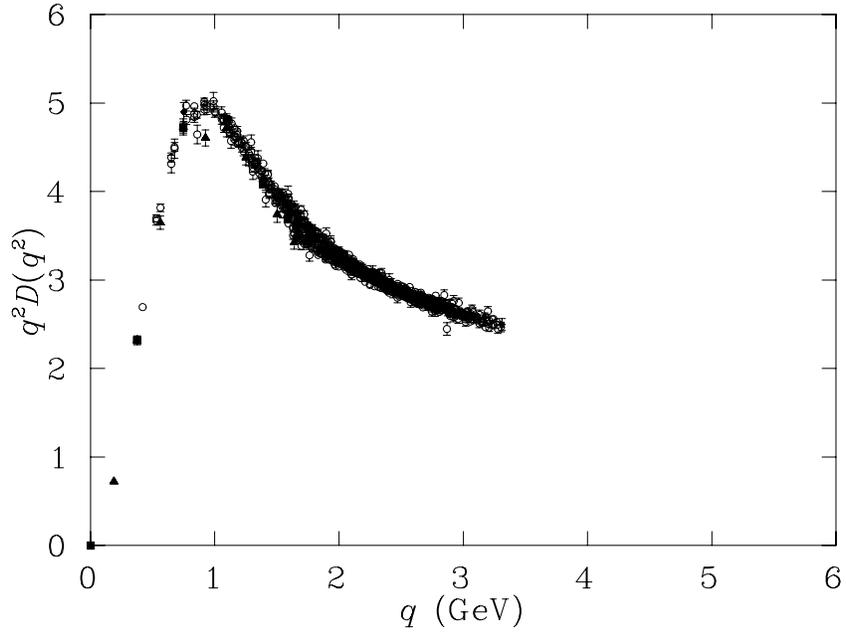,angle=90,width=11cm}
\end{center}
\caption{Gluon propagator from lattice 5 at $\beta = 4.10$, which has
spacing $a\simeq 0.27 \text{ fm}$ on $12^3 \times 24$.  This has the
same physical volume as lattice 3 of
Fig.~\protect\ref{fig:b375PhysAll}.  The propagator is shown for all
momenta (no data cuts) after tree-level correction.}
\label{fig:b410PhysAll}
\end{figure}

\begin{figure}[p]
\begin{center}
\epsfig{figure=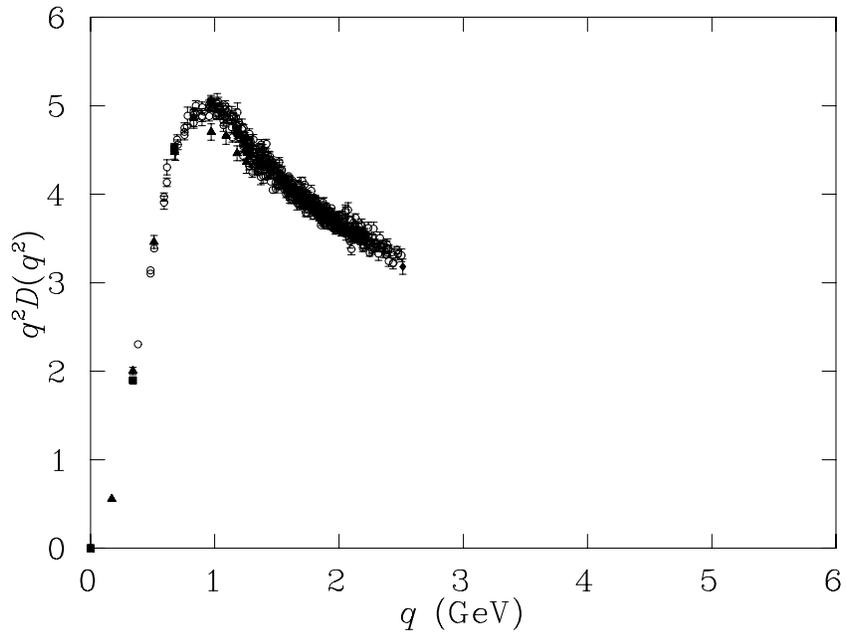,angle=90,width=11cm}
\end{center}
\caption{Gluon propagator from lattice 2, the smaller lattice at
$\beta=3.92$ which has spacing $a\simeq 0.35 \text{ fm}$ on a $10^3
\times 20$ lattice.  Finite volume errors are just detectable as
indicated by momenta along the time axis (filled triangles) falling
below the rest of the data.  Tree-level correction has been used, but
no data cuts have been applied.}
\label{fig:b392sPhysAll}
\end{figure}

\begin{figure}[p]
\begin{center}
\epsfig{figure=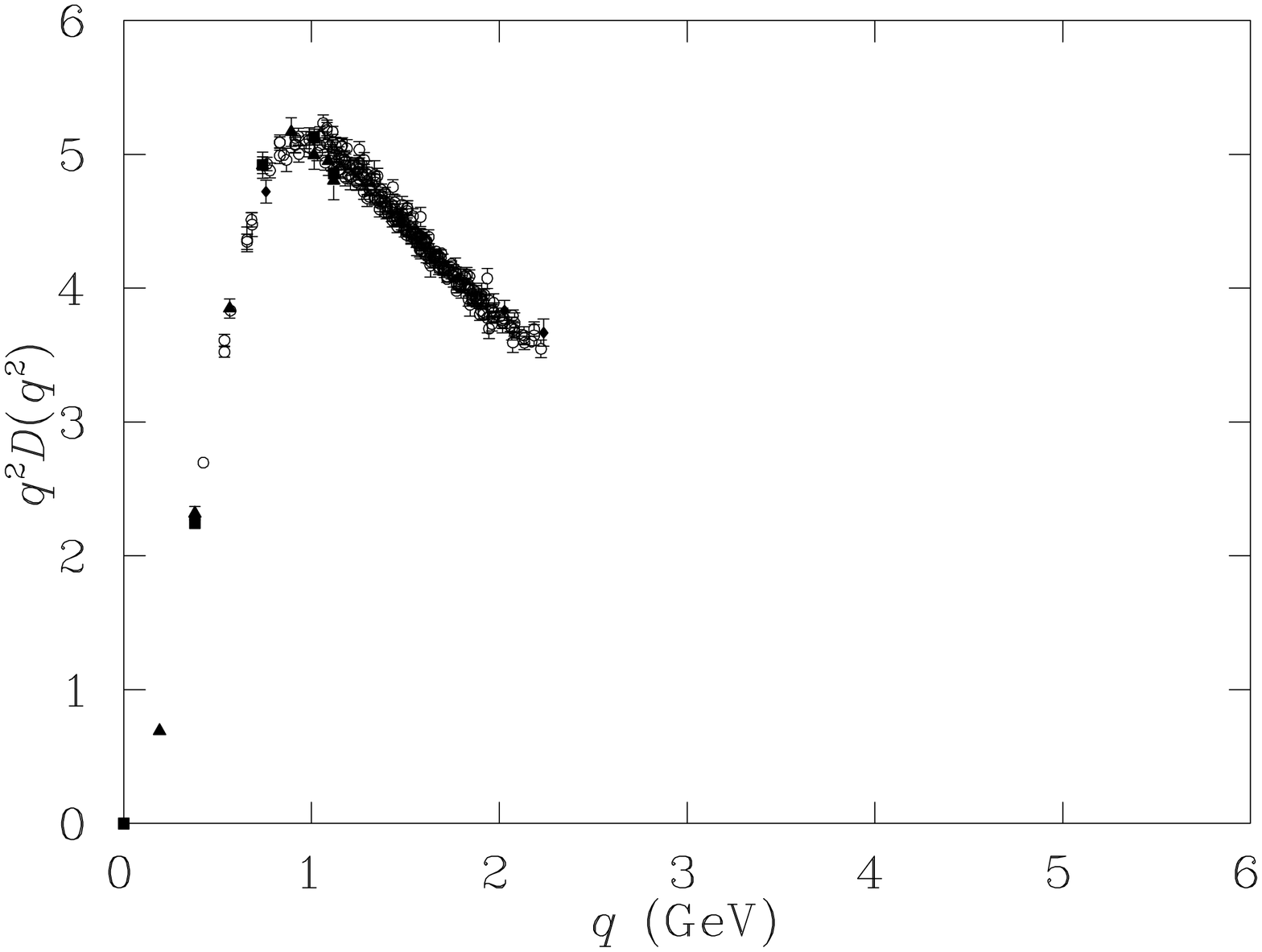,angle=90,width=11cm}
\end{center}
\caption{Gluon propagator from lattice 3 at $\beta = 3.75$, which has
spacing $a\simeq 0.41 \text{ fm}$ on $8^3\times 16$.  The propagator
is shown for all momenta (no data cuts) after tree-level correction.
This propagator is consistent with that obtained on much finer
lattices.}
\label{fig:b375PhysAll}
\end{figure}

\subsection{Lattice Spacing Dependence}

At this point it is interesting to explore the effect of making the
lattice coarser.
Figures~\ref{fig:b410PhysAll},~\ref{fig:b392sPhysAll}
and~\ref{fig:b375PhysAll} show the uncut, tree-level corrected
propagator on progressively coarser lattices ($a =$ 0.27, 0.35 and
0.41 fm respectively).  Consider the most extreme case, shown in
Fig.~\ref{fig:b375PhysAll}.  This very coarse lattice has spacing
$a=0.41\text{ fm}$, which is more than twice as coarse as the previous
lattices.  Any sign of a perturbative tail has been lost, as the UV
cutoff has been lowered, but the infrared behavior remains.  There is
no sign of any qualitative change, which appears to indicate that even
on such a coarse lattice we are not losing information vital to the
infrared physics of the gluon propagator.  

\begin{figure}[p]
\begin{center}
\epsfig{figure=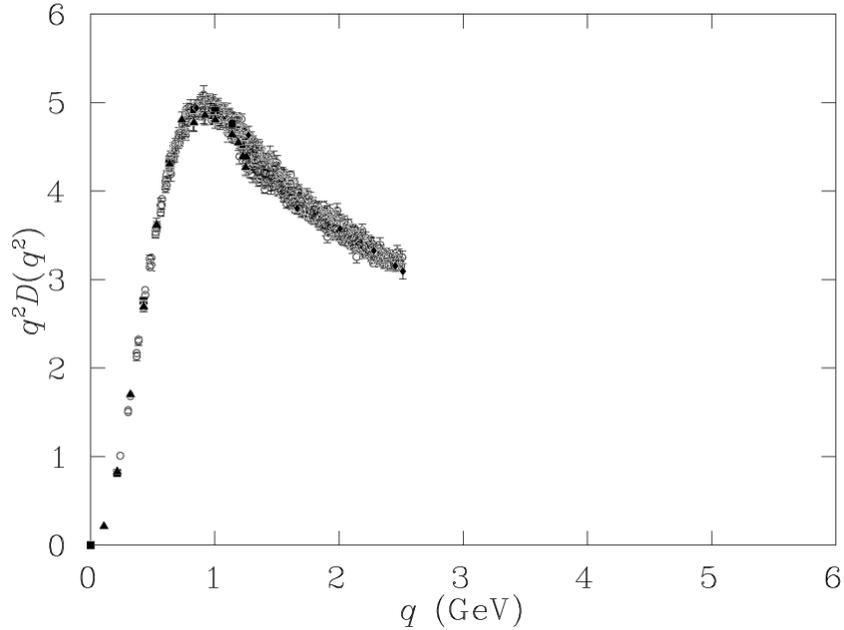,angle=90,width=11cm}
\end{center}
\caption{Gluon propagator from lattice 4, the larger lattice at
$\beta=3.92$, which has spacing $a\simeq 0.35 \text{ fm}$ on a $16^3
\times 32$ lattice providing the largest physical volume of any in
this study.  Tree-level correction has been used, but no data cuts
have been applied.}
\label{fig:b392lPhysAll}
\end{figure}

This gives us great confidence in the use of improved actions on
coarse lattices for the probing of nonperturbative physics.  This is
the motivation for creating lattice 4 at $a = 0.35$ on a very large
volume.  Fig~\ref{fig:b392lPhysAll}, which shows the results from this
large lattice, shows no signs of significant finite volume artifacts
when compared with Fig.~\ref{fig:b392sPhysAll} which has the same
lattice spacing, but a smaller volume.

\subsection{Data Cuts}

Having identified possible lattice artifacts, cuts may be applied to clean up
the data, making it easier to draw conclusions about continuum physics.  Data
at large momenta will of course be most susceptible to finite lattice spacing 
errors.
We choose to prefer data from momentum points near the four-diagonal, as this 
evenly samples all Cartesian
directions, i.e., for a given momentum squared ($q^2$) it 
has the smallest values of each of the Cartesian components $q_\mu$.  This 
should minimize finite lattice spacing artifacts. 

We calculate the distance 
$\Delta\qhat$ of a momentum vector $\qhat$ from the diagonal using
\begin{equation}
\Delta\qhat = |\qhat|\sin\theta_{\qhat},
\end{equation}
where the angle $\theta_{\qhat}$ is given by
\begin{equation}
\cos\theta_{\qhat} = \frac{\qhat \cdot \hat{n}}{|\qhat|},
\end{equation}
and $\hat{n} = \frac{1}{2} (1,1,1,1)$ is the unit vector along the
diagonal.  In this way we ignore data points that are potentially most
affected by hypercubic artifacts.  We call this cut the \emph{cylinder
cut}~\cite{Lei98}.  From this point on, we exclude points greater than
two spatial momentum units\footnote{A spatial momentum unit is $2\pi /
a L_s$ where $L_s$ is the number of lattice sites in the
spatial directions ($L_s = L_x = L_y = L_z$).  } from the
four-diagonal.  Furthermore, the point at zero four-momentum has been
cut from all the following plots of $q^2D(q)$.  On any finite lattice,
$D(0)$ must be finite, hence $q^2 D(q) = 0$ for $q = 0$.  This point
is therefore trivial when plotting $q^2 D(q)$.  When the scalar
function, $D(q)$, itself is considered we can make a study of $D(0)$
by considering it on lattices of differing volumes and then making an
infinite volume extrapolation.  We will perform this extrapolation
below.

\subsection{Action Dependence}

\begin{figure}[p]
\begin{center}
\epsfig{figure=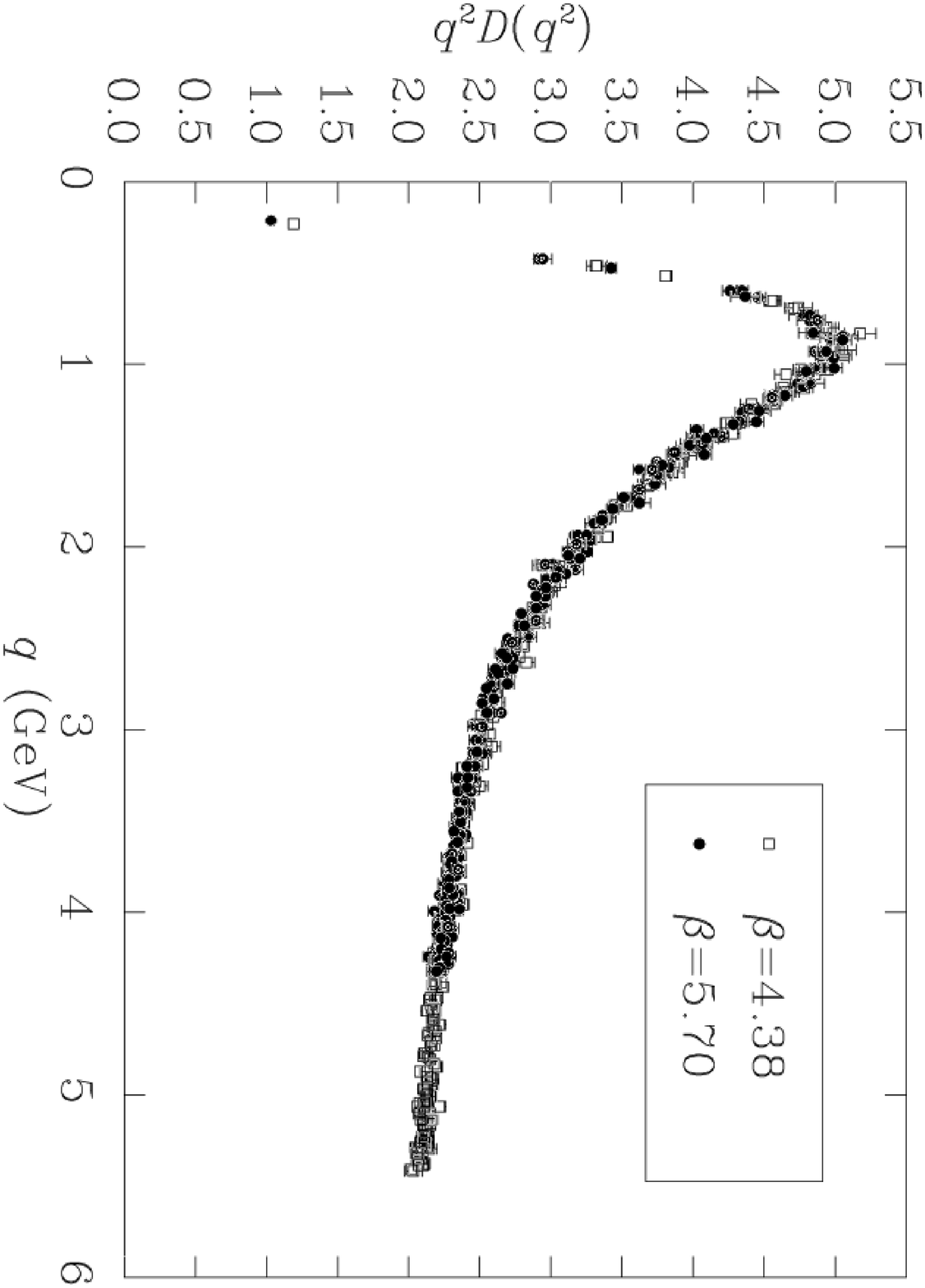,angle=90,width=11cm}
\end{center}
\caption{Comparison of the gluon propagator from lattices 1w at $\beta
= 5.70$ and 1i at $\beta = 4.38$.  Data has been cylinder cut and tree-level 
correction has been applied.
We have determined $Z_3(\text{improved}) / Z_3(\text{Wilson}) = 1.09$
by matching the vertical scales of the data.}
\label{fig:Comp1w_1i}
\end{figure}

Once again we compare the gluon propagator generated with the Wilson
action to that generated with the improved action after tree-level
correction, this time applying the cylinder cut.  To make the
comparison in Fig.~\ref{fig:Comp1w_1i}, we note that there is of
course a small difference in normalization.  This is the difference in
the $Z_3$ renormalization between the Wilson and improved propagators.
As the relative renormalization is $q^2$ independent, the unimproved
propagator has been multiplied by a relative renormalization of 1.09
to make direct comparison possible.  This factor is deduced by
adjusting the vertical scales of the two data sets until they agreed.
Apart from the superior performance of the improved propagator, which
has already been discussed, the two actions produce the same result.

\begin{figure}[p]
\begin{center}
\epsfig{figure=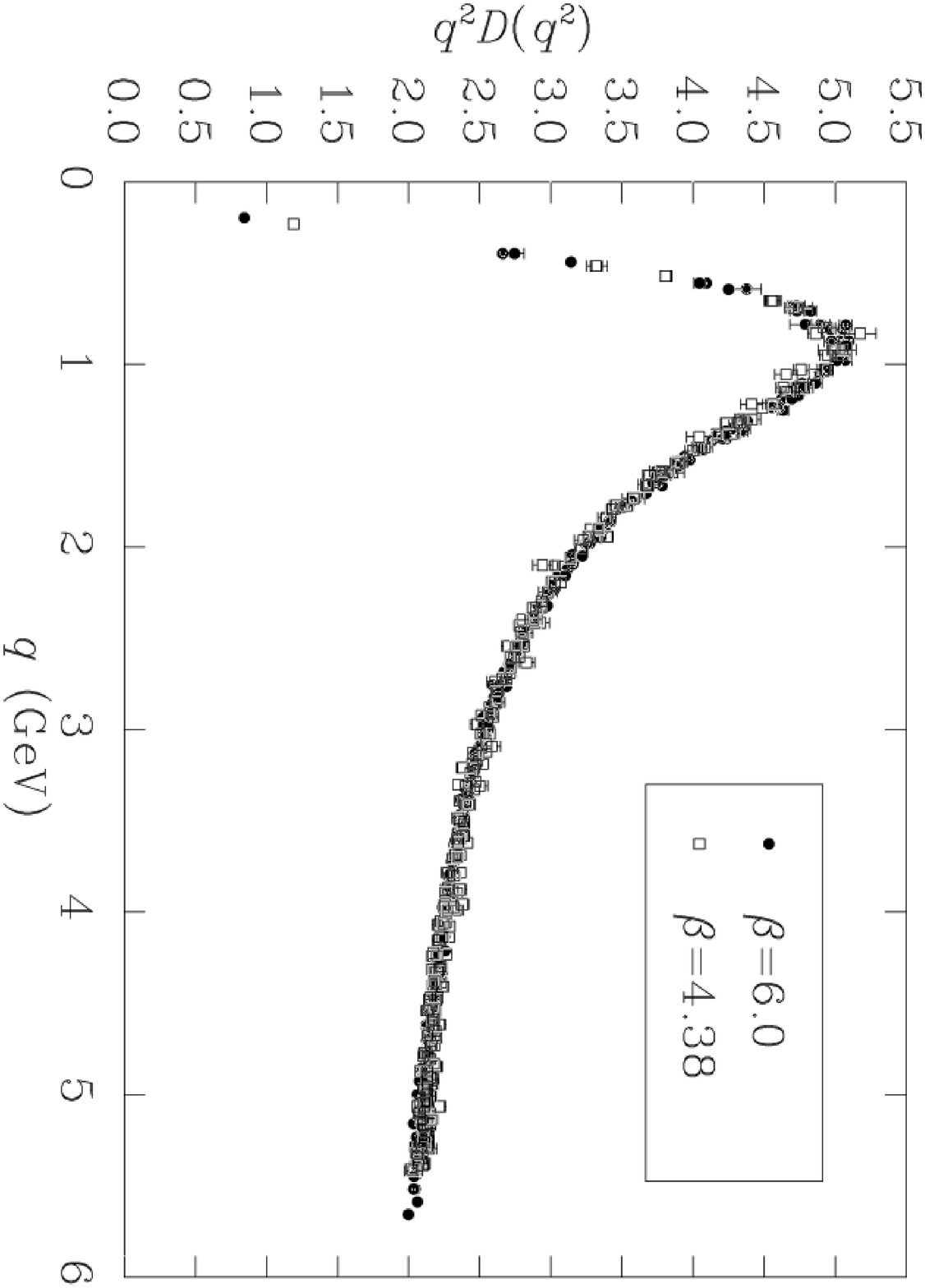,angle=90,width=11cm}
\end{center}
\caption{Comparison of the gluon propagator from the finest improved
lattice (lattice 1i, $\beta = 4.38$) and the finest Wilson lattice 
(lattice 6, $\beta=6.0$).  Data has been cylinder cut and the appropriate
tree-level corrections have been applied.  The data from lattice 6 is
half-cut whereas lattice 1i displays the full Brillouin zone.  We have
determined $Z_3(\text{improved}) / Z_3(\text{Wilson}) = 1.08$ by
matching the vertical scales of the data.}  
\label{fig:Comp1i_6}
\end{figure}

We push our results further by comparing the improved $\beta=4.38$
propagator with that from lattice 6 (Wilson action), which is finer
($a=0.1\text{ fm}$), has more points ($32^3\times 64$) and is a little
larger.  Both data sets are cylinder cut, and each is tree-level
corrected according to its action.  The relative renormalization has
been determined to be $Z_3(\text{improved}) / Z_3(\text{Wilson}) =
1.08$.  It can be seen from Fig.~\ref{fig:Comp1i_6} that not only are
the two propagators consistent, but that the ultraviolet performance
of lattice 1i is remarkable.  The propagator from Ref.~\cite{Lei98}
had the momentum half-cut applied, whereas our improved propagator
with lattice spacing $a=0.17\text{ fm}$ is shown for the entire
Brillouin zone.  We have calculated the propagator over the same range
of momenta as Ref.~\cite{Lei98}, despite using a much coarser lattice.

\subsection{Scaling Analysis}

\begin{figure}[p]
\begin{center}
\epsfig{figure=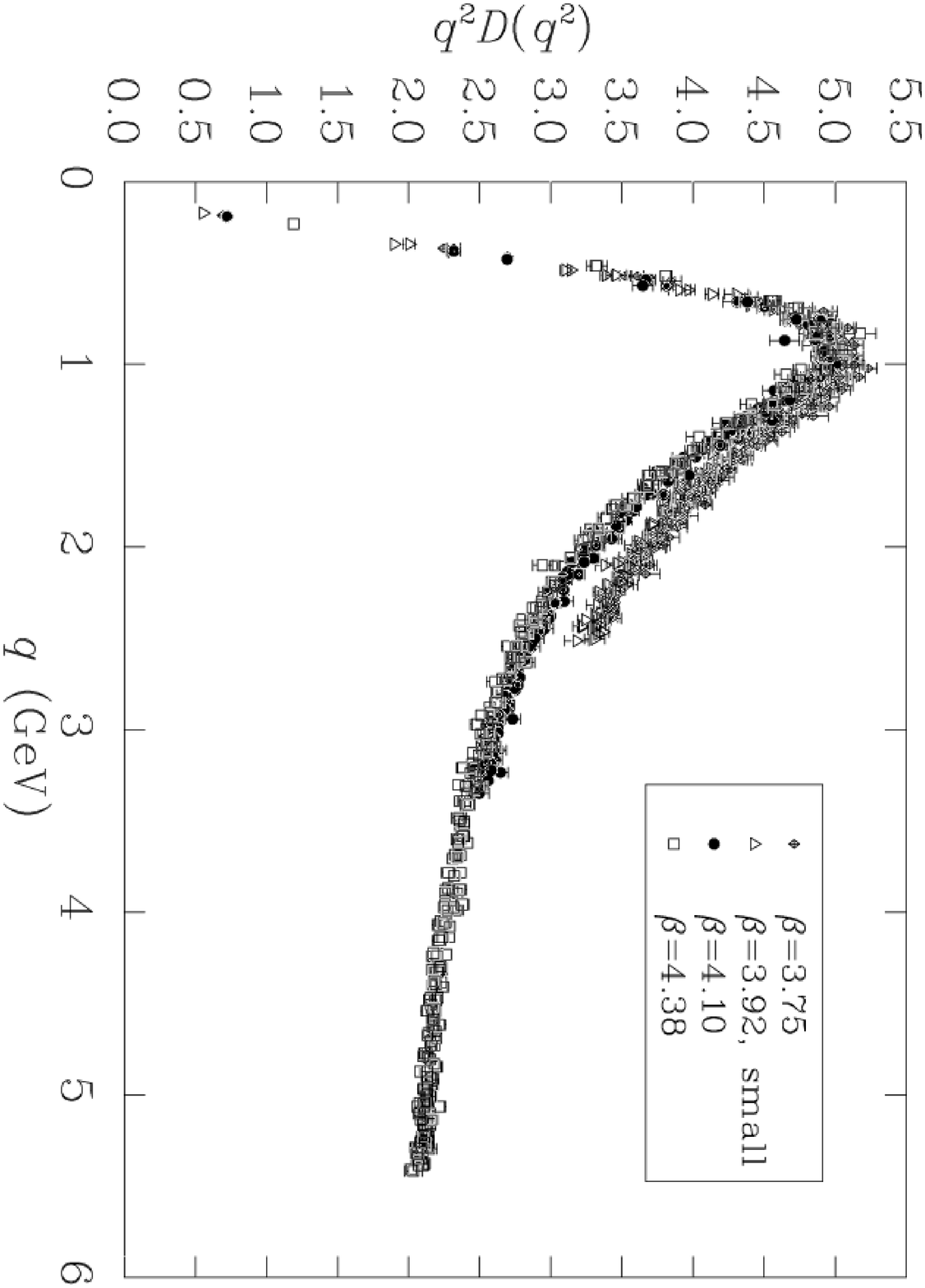,angle=90,width=11cm}
\end{center}
\caption{Comparison of the gluon propagator from lattices 1i ($\beta=4.38$),
2 ($\beta=3.92$, small), 3 ($\beta=3.75$), and 5 ($\beta=4.10$), 
which have a variety of lattice spacings.  Data has been cylinder cut and 
tree-level correction has been applied.  Data 
from the two finest improved lattices (0.17 and 0.27 fm) are consistent.  A 
clear violation of scaling is seen in the coarsest two lattices (0.35 and 
0.41~fm), where the spacing is too coarse for tree-level 
correction to completely restore the full Brillouin zone behavior.}
\label{fig:Comp_betas}
\end{figure}

Next, we consider the propagator on the coarser lattices.
Fig.~\ref{fig:Comp_betas} shows the propagator from lattices 1i, 2, 3
and 5.  Examining Figures~\ref{fig:Comp1w_1i} and~\ref{fig:Comp_betas}
we see that the Wilson $\beta=5.7$ and improved $\beta=4.10$ and
$\beta=4.38$ results all agree well, which suggests that these are
``fine enough'' lattices.  We see that the $\beta=3.75$ and
$\beta=3.92$ propagators do not quite line up with the others, but
instead the UV tail rises slightly as the lattice becomes coarser.
This is an indication of a loss of scaling.  The lattices at
$\beta=3.92$ and $\beta=3.75$ having $a = 0.35$ and 0.41 fm
respectively are too coarse for the tree-level correction to
completely correct the entire Brillouin zone, which is not surprising.
We have placed extraordinary demands on our simulations by examining
them near the cutoff.  The conclusion is that such coarse lattices
should be half-cut.  Nevertheless, the propagators all agree in the
infrared. 
Now that we have an understanding of the dependence of
lattice propagator on the lattice spacing, we can study the effect of
the finite volume.

\begin{figure}[p]
\begin{center}
\epsfig{figure=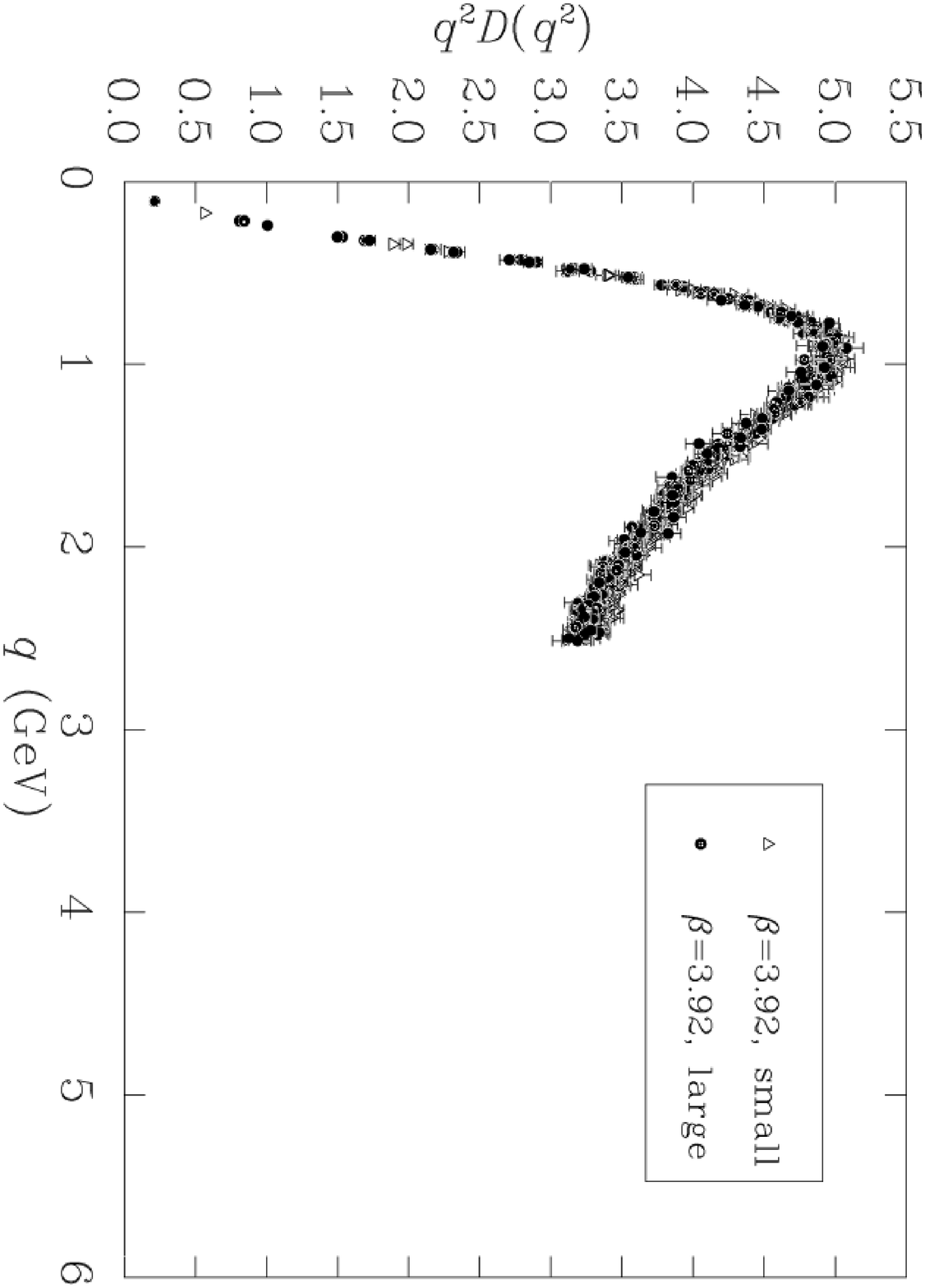,angle=90,width=11cm}
\end{center}
\caption{Comparison of the gluon propagator from lattices 2 and 4,
which have the same lattice spacing ($a=0.35 \text{ fm}$), but
different lattice volumes.  Notice that one lies directly over the
other, despite having very different volumes.  Data has been cylinder
cut and tree-level correction has been applied.}
\label{fig:Comp2_4}
\end{figure}

\subsection{Volume Dependence}

Results from lattices 2 and 4 have already been reported in 
Ref.~\cite{Bon00} and are presented again here for completeness
and ease of comparison.
They have same lattice spacing, but different numbers of 
lattice points, and hence different physical volumes.  The gluon propagator 
has been calculated
on each lattice, and the results compared in Fig.~\ref{fig:Comp2_4}.  The two
propagators are consistent in this figure, despite the fact that one
lattice has
sides 60\% longer in all four directions.  This shows that finite volume 
effects are small compared to the statistical errors.  The turn over 
seen in the gluon propagator in lattice studies is certainly not a finite
volume effect.  Note that $5.65^3 \times 11.30 \text{ fm}^4$ is a very large 
volume by the
standards of present day lattice studies, and gives us an unprecedented
look at the behavior of QCD in the deep infrared.

\begin{figure}[t]
\begin{center}
\epsfig{figure=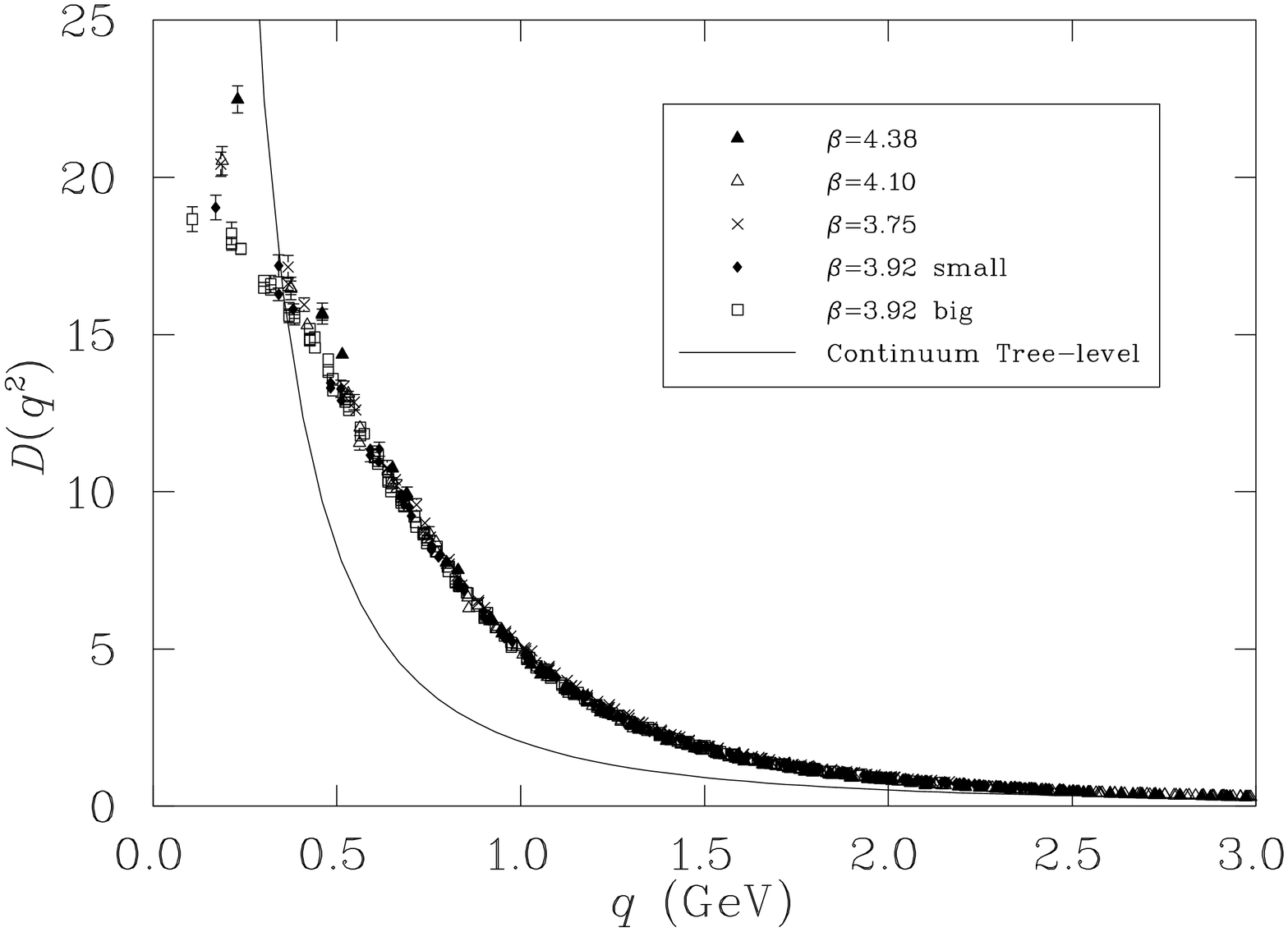,angle=90,width=11cm}
\end{center}
\caption{Comparison of the gluon propagator generated with an improved action
on five different lattices.  We find good agreement down to 
$q \simeq 500$ MeV.  At the lowest accessible momenta the data points drop
monotonically with increasing volume, but the lowest point (on the largest 
lattice) shows signs of having converged to its infinite volume value.
For comparison with perturbation theory, a plot of the continuum, tree-level 
gluon propagator (i.e., $1 / q^2$ appropriately scaled) has been included.}
\label{fig:AllProps}
\end{figure}

\begin{figure}[t]
\begin{center}
\epsfig{figure=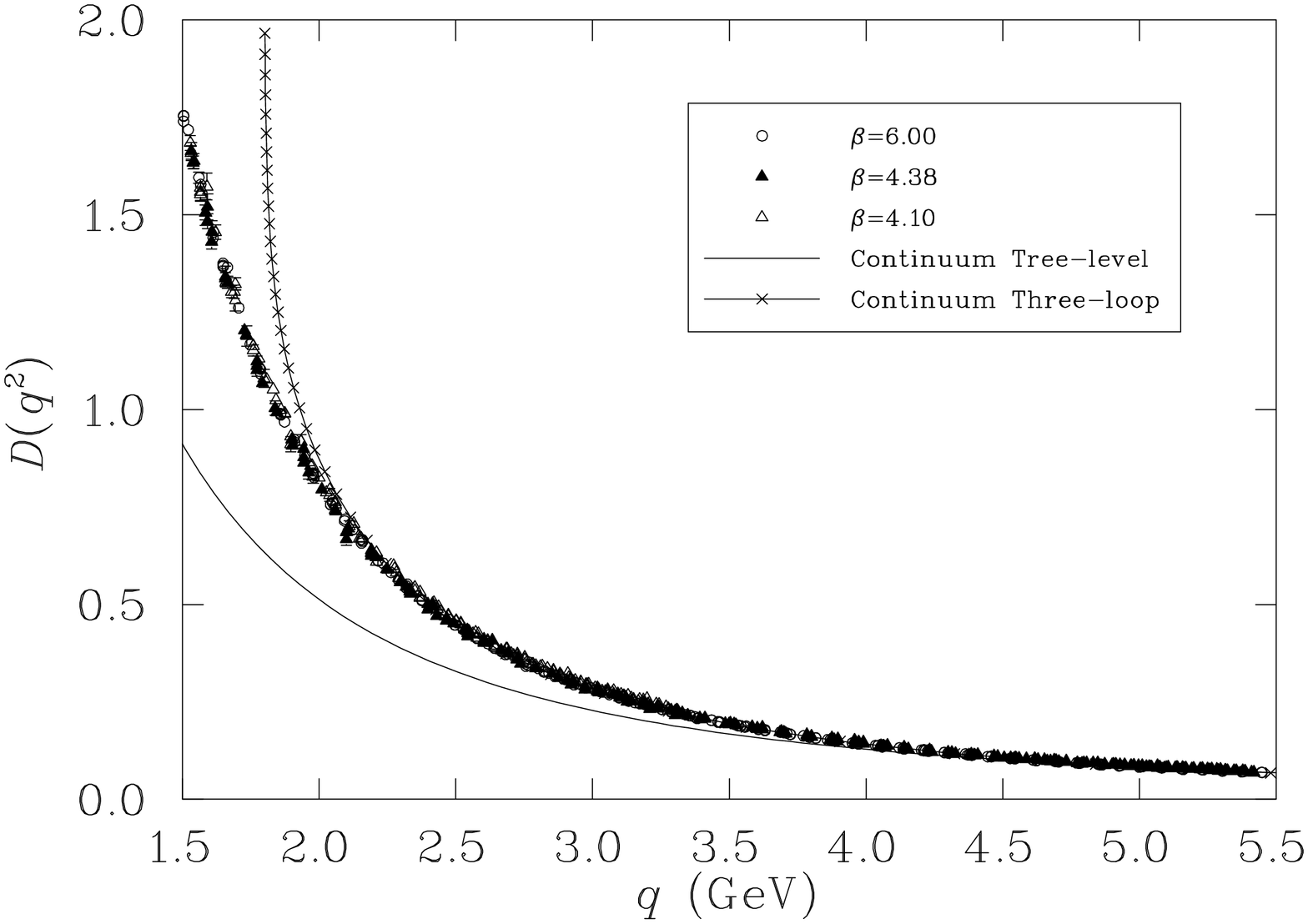,angle=90,width=11cm}
\end{center}
\caption{Comparison of the lattice gluon propagator with that obtained from
perturbation theory, in the ultraviolet to intermediate regime.  The continuum
expressions are tree-level (i.e., $1 / q^2$ appropriately scaled) and
the three-loop expression used in Ref.~[2].}
\label{fig:pert_vs_latt}
\end{figure}

Fig.~\ref{fig:AllProps} shows the cylinder-cut
data for the scalar function $D(q^2)$ 
for each of the improved lattices.  This plot provides a dramatic 
demonstration of 
lattice artifacts.  In this way of plotting our results, the five
lattices appear in very good agreement
in the ultraviolet and through intermediate momenta.  
When plotted in this way, we
can see that below $\sim 500 \text{ MeV}$ the 
propagators do begin to differ due to finite volume effects.  As the volume
increases, the low momenta data points drop, until we can see the infrared
flatten off.  The grouping of points around 400 MeV suggest that we have,
for the two largest lattices, results indicative of the infinite volume limit.
At $\sim 250 \text{ MeV}$, the results for the two largest lattices 
(both $\beta=3.92$) are 
consistent, and in particular the fact that the small difference between them
is produced by such a large difference in volume gives us confidence in
the results.  For comparison, the tree-level, perturbative expression 
$D(q^2) = 1 / q^2$ is also shown, suitably normalized.

It is interesting to note that the disagreement in the propagators
above 1~GeV or so revealed in Fig.~\ref{fig:Comp_betas} is hidden by
the scale of the vertical axis in Fig.~\ref{fig:AllProps}.  Multiplication
of the propagator by $q^2$ is required to amplify this region and
critically examine the extent to which lattice spacing artifacts are
removed by ${\cal O}(a^2)$ improvement terms.  A failure to do this
could lead to incorrect conclusions being drawn on the effectiveness
of improvement in the gluon propagator.  Thus it is always best to
plot $q^2D(q^2)$ versus $q^2$, when the hypercubic artifacts are
of interest.

\subsection{Asymptotic Behavior}

For further comparison with perturbation theory, we have chosen to show the 
gluon propagator from 1.5 to 5.5 GeV, in Fig.~\ref{fig:pert_vs_latt}.  In this
window, the transition from perturbative to nonperturbative physics can be 
clearly seen.  As well as the lattice gluon propagator and the tree-level, 
continuum propagator, we show a perturbative, three-loop 
calculation~\cite{Boucaud}.  We used parameters obtained from 
Ref.~\cite{Bec99}, where at the renormalization point, $\mu = 5.48$ GeV, the
strong coupling constant was found to be $\alpha(\mu) = 0.255$.  That was a
quenched calculation, so this number should not be compared directly with 
experiment.  The data agree very well with three-loop perturbation theory down
to $q\simeq 2.5$~GeV.  Below 2~GeV we see that three-loop perturbation
theory begins to fail.

\subsection{Propagator at Zero Four-Momentum}

\begin{figure}[t]
\begin{center}
\epsfig{figure=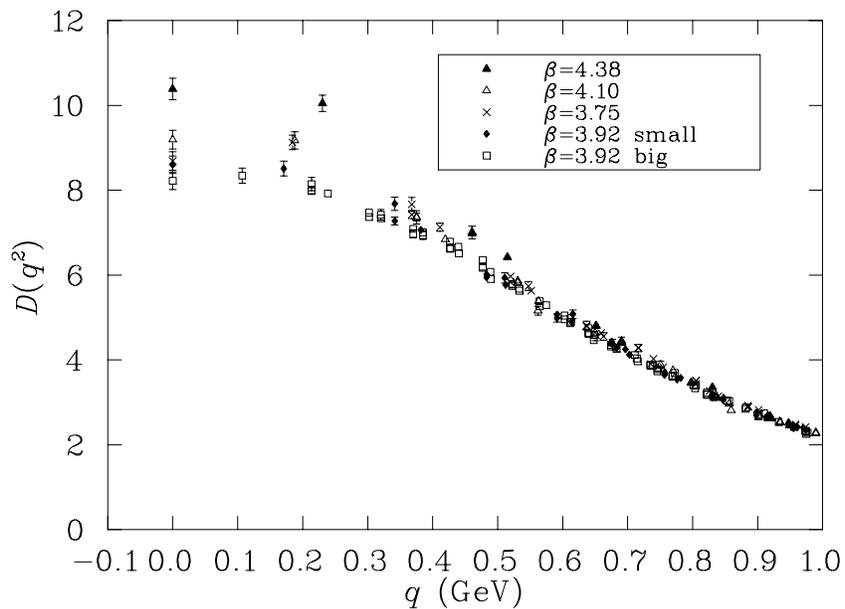,angle=90,width=11cm}
\end{center}
\caption{The renormalized gluon propagator is shown in the infrared region,
including the points at zero four-momentum, from five lattices.}
\label{fig:infrared}
\end{figure}

\begin{figure}[t]
\begin{center}
\epsfig{figure=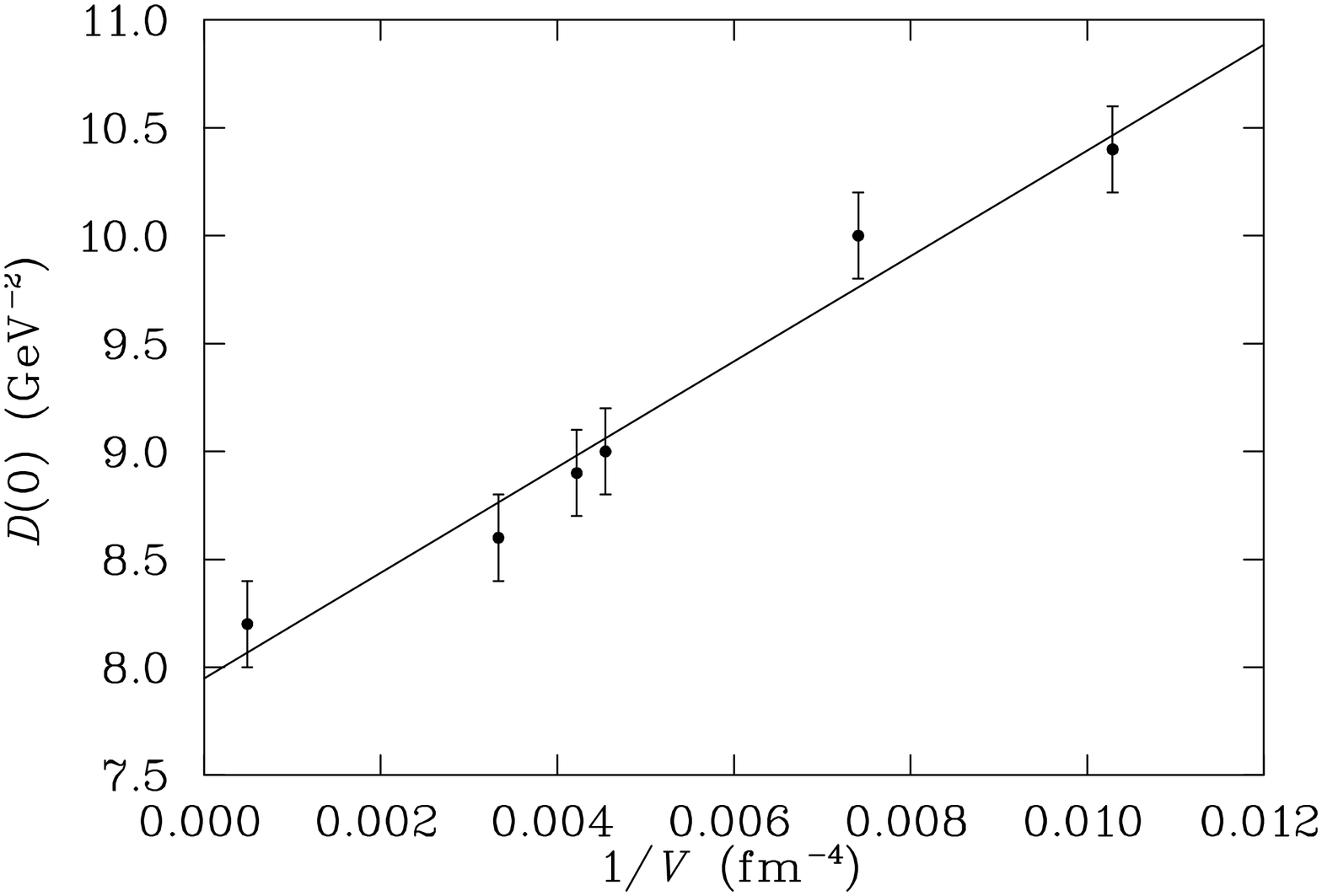,angle=90,width=11cm}
\end{center}
\caption{Values for the gluon propagator at zero four-momentum,
$D(0)$, plotted as a function of the inverse lattice volume.  The
solid line represents a linear fit to the lattice results.  The fit
indicates the largest volume results are very close to the infinite
volume limit and $D(0) = 7.95(13)$ GeV${}^{-2}$ in the infinite volume
limit.}
\label{fig:volExtrap}
\end{figure}

Values for the gluon propagator at zero four-momentum are shown in
Table~II for each of the lattices created in this investigation.
Statistical errors are given in parentheses.
The renormalization
condition of Eq.~(\ref{eq:b.c.}) is enforced at the renormalization
point $\mu = 4.0$ GeV, which sets the scale for $D(q^2)$.
We see that as the volume of the lattice
increases, $D(0)$ becomes smaller.  
In Fig.~\ref{fig:infrared} we plot the infrared behavior of the
renormalized gluon propagator for five lattices and we include the
points calculated at zero momentum in this plot.  We see that the
the infrared behavior is quite smooth and reasonably consistent
for our two largest volume lattices ($\beta=3.92$, small and large).
Fig.~\ref{fig:volExtrap}
illustrates the data with a linear fit in the inverse volume according
to
\beq
D(0) = c {1 \over V} + D_\infty(0)
\eeq
We find a reasonable fit with parameter values $c = 245(22)\ {\rm fm}^4 \,
{\rm GeV}^{-2}$ and $D_\infty(0) = 7.95(13)$ GeV${}^{-2}$, where
$D_\infty(0)$ is the infinite volume limit of the zero-momentum
gluon propagator.  Fig.~\ref{fig:volExtrap} strongly
supports the hypothesis that the gluon propagator is finite in the
infrared.  It is also clear that the results of our largest physical
volume lattice are very close to the infinite volume limit.

\begin{table}
\label{tab:Dzero}
\centering
\begin{tabular}{cccccc}
Lattice & Dimensions      & $\beta$ &    D(0)   & D(0) ($\text{GeV}^{-2}$) & Volume ($\text{fm}^4$)  \\
\hline
  1i    & $16^3\times 32$ &   4.38  & 32.0 (8)  & 10.4 (2) &  97.2 \\
  1w    & $16^3\times 32$ &   5.70  & 24.0 (5)  & 10.0 (2) & 135   \\ 
  5     & $12^3\times 24$ &   4.10  & 10.6 (3)  &  9.0 (2) & 220   \\
  3     & $8^3 \times 16$ &   3.75  &  4.3 (1)  &  8.9 (2) & 237   \\
  2     & $10^3\times 20$ &   3.92  &  5.7 (1)  &  8.6 (2) & 300   \\
  4     & $16^3\times 32$ &   3.92  &  5.4 (1)  &  8.2 (2) & 2038   \\
\end{tabular}
\caption{The value of gluon propagator at zero
four-momentum for each of the lattices created in this investigation,
in order of increasing volume.  The raw (dimensionless) and physical
values are given.  In obtaining the physical values we have set the
renormalization condition $D(\mu^2) = 1 / \mu^2$ at $\mu = 4.0$ GeV.
An estimate of the uncertainty in the last figure is given in
parentheses.}
\end{table}

Note that a complete systematic extrapolation to the infinite volume
limit remains to be carried out in the future.  Ideally, one
performs a number of calculations at fixed volume and various lattice
spacings and then performs a continuum limit extrapolation for that
fixed volume.  This continuum limit extrapolation would be done for
each of a variety of lattice volumes and then finally an infinite
volume extrapolation performed on those results.  This procedure
corresponds to the axiomatic field theory prescription of taking the
continuum limit before the infinite volume limit.  Given this caution,
the finite precision of this study, and the fact that the linear
ansatz above may be incorrect, we can not completely exclude the
possibility that the deep infrared (i.e., below $\sim$ 350 MeV)
behavior of $D(q^2)$ may very slowly decrease toward zero as the
infinite volume limit is taken.  

It is interesting to compare our results with a recent calculation of
the gluon propagator in Laplacian gauge~\cite{lapprop}, which is
expected to be free of gauge ambiguity.  In that gauge, the propagator
takes its perturbative, Landau-gauge value in the asymptotic region and
is also infrared finite.  The Laplacian gauge propagator is seen to
have a behavior similar to that seen here.

\section{Conclusions}

The gluon propagator has been calculated on a set of lattices with an
\oa{2} mean-field improved action, in \oa{2} mean-field improved
Landau gauge.  Tree-level correction has been shown to reduce
rotational symmetry breaking and dramatically improve the ultraviolet
behavior of the propagator.

For $\beta \geq 4.10$ ($a \leq 0.27$ fm) the tree-level corrected improved
propagator displays scaling over the entire Brillouin zone.  At $\beta
= 4.38$ ($a = 0.166$ fm), the gluon propagator has excellent behavior
for the entire range of available momenta in the Brillouin zone,
reproducing the anticipated UV behavior of perturbation theory to
three-loops.

The infrared behavior of the gluon propagator is robust
even with a lattice spacing of 0.41 fm.  
Calculation on a lattice with a large volume indicates that finite
volume effects are small.  In particular, the turn over observed in
previous studies of the Landau gauge gluon propagator is not a finite
volume artifact.  We conclude that the propagator is almost certainly
infrared finite, in agreement with earlier studies.
A significant volume dependence is revealed only at the smallest
non-trivial momenta.  An extrapolation of $D(0)$ via a linear ansatz
inversely proportional to the physical lattice volume provides a
reasonable fit.  Moreover, results from our largest volume lattice
reside very close to the infinite volume limit.  
We have probed the approach to the infinite volume limit by first
determining a range of $\beta$ in which the propagator scales for $q <
0.7$ GeV on similar finite volumes.  Physically large volumes are
accessed by decreasing $\beta$ within the scaling range on large
lattices.  A more complete study of the infinite volume
limit should be undertaken in the near future.

The tree-level corrected results from our $\beta = 3.92$ ($a =
0.353$ fm) $16^3 \times 32$ lattice with a physical volume of $5.65^3
\times 11.30 = 2038$ fm${}^4$ may be regarded as an excellent estimate
of the infinite volume, continuum limit Landau-gauge gluon
propagator for $q < 0.7$ GeV.  The tree-level corrected results from our
$\beta = 4.38$, ($a = 0.166$ fm) results presented here are an
excellent estimate of the infinite volume, continuum limit of the
Landau-gauge gluon propagator for $q > 0.7$ GeV.
We have seen that these two sets
of data smoothly match in the intermediate regime ($q \sim 0.7$ GeV)
and are entirely consistent with each other in this region.
The possible effects of lattice Gribov copies remains a very
interesting question and we plan to extend this study to Laplacian gauge
and other related gauge-fixing schemes in the near future.

\section*{Acknowledgments}

POB would like to acknowledge helpful conversations with Urs Heller and 
correspondence with Phillippe Boucaud.  This work was supported by the 
Australian Research Council and by grants of supercomputer time on the CM-5 
made available through the South Australian Centre for Parallel Computing. 
The work of POB was supported in part by DOE contract DE-FG02-97ER41022.

%%%%%%%%%%%%%%%%% Bibliography %%%%%%%%%%%%%%%

\end{document}